\documentclass[aps,pra,twocolumn,groupedaddress,byrevtex]{revtex4}

\usepackage{graphicx}
\newcommand{\sgmsnl}{\tilde\sigma_{s-i}^2}
\newcommand{\sgmsnlsat}{\tilde\sigma_{s-i,{\rm sat}}^2}
\newcommand{\Dx}{\Delta x_{\rm symm}}
\begin{document}

\title{Quantum spatial correlations in high-gain parametric down-conversion measured by means of a CCD camera}

\author{O. Jedrkiewicz, E. Brambilla, M. Bache, A. Gatti,
L. A. Lugiato, and P. Di Trapani}

\affiliation{INFM, Dipartimento di Fisica e Matematica,
Universit{\`a} dell'Insubria, Via Valleggio 11, 22100 Como, Italy.}

\date{\today}

\begin{abstract}
We consider travelling-wave parametric down-conversion in the
high-gain regime and present the experimental demonstration of the
quantum character of the spatial fluctuations in the system. In
addition to showing the presence of sub-shot noise fluctuations in
the intensity difference, we demonstrate that the peak value of
the normalized spatial correlations between signal and idler lies
well above the line marking the boundary between the classical and
the quantum domain. This effect is equivalent to the apparent
violation of the Cauchy-Schwartz inequality, predicted by some of
us years ago, which represents a spatial analogue of photon
antibunching in time. Finally, we analyse numerically the
transition from the quantum to the classical regime when the gain
is increased and we emphasize the role of the inaccuracy in the
determination of the symmetry center of the signal/idler pattern
in the far-field plane.
\end{abstract}

\maketitle

\section{Introduction}
Quantum noise reduction has been the subject of extensive research
in the past. Although most of the experimental effort has been
concentrated on the generation and applications of
quadrature-squeezed light, amplitude-squeezed light and intensity
correlated twin beams have also been produced. The possibility of
quantum correlations between light beams has played a major role
in the recent development of quantum optics. In particular the
spatial aspects of correlations of quantum optical fluctuations
have been the object of several studies (see for example
\cite{Kolobov99, Lugiato99, Pittman96, Burkalov97, Saleh00,
Gatti95}) and have several new and promising applications such as
quantum holography \cite{abouraddy01}, the quantum teleportation
of optical images \cite{Sokolov01}, and the measurements of small
displacements beyond the Rayleigh limit \cite{Treps02}. An
overview of this relatively new branch of quantum optics, defined
as quantum imaging, can be found in \cite{Lugiato02}.

The process of parametric down-conversion (PDC) is particularly
suitable for the study of spatial effects, because of its large
emission bandwidth in the spatial frequency domain
\cite{Devaux00}. There is now a large literature on spatial
effects in the low-gain regime, where photon pairs are detected
via coincidence counting \cite{Lugiato02}. Nevertheless, to date
spatial correlation measurements in this regime have not evidenced
any relevant quantum effect \cite{Jost98, Oemrawsingh02}. On the
other side, measurements performed in the medium-gain regime (pump
power ${\leq1}$ MW) evidenced the twin beam character of the PDC
emission \cite{Aytur90}, i.e. a sub-shot-noise correlation between
the photon numbers of the whole signal and idler beams. Recent
theoretical investigations done for an arbitrary gain
\cite{Brambilla04, Gatti99} have predicted multi-mode spatial
correlations below shot-noise between several portions of the
signal and idler emission cones that correspond to phase conjugate
modes. There is a minimum size of the modes that have to be
detected in order to observe a quantum correlation, which we shall
refer to as the {\em coherence area}. This is determined by the
conditional uncertainty in the directions of propagation of the
twin photons, and is roughly given by the inverse of the
near-field gain area \cite{Brambilla04}.

In this paper we present a description of the far field detection
of the PDC radiation emitted by a $\beta$-barium borate (BBO)
non-linear crystal pumped by a low-repetition rate (2 Hz) pulsed
high-power laser (1GW-1ps). The detection is performed by means of
a high quantum efficiency ($\eta\approx89\%$) scientific CCD
camera. In a recent letter \cite{Jedrkiewicz04} we presented
measurements showing the existence of quantum spatial correlation
between the signal and idler beams. The aim of this paper is, on
the one side, to describe in details several features of the
experiment (such as the asymmetry of the PDC beam distribution,
the CCD diagnostic) which were not presented in
\cite{Jedrkiewicz04}. On the other side, we shall provide a
careful numerical analysis and interpretation of the experimental
data, which (i) will enlighten a different aspect of the quantum
noise correlation -namely an apparent violation of the
Cauchy-Schwarz inequality-, and (ii) will discuss the role of the
finite size of the detection pixel and of the inaccuracy in the
determination of the symmetry center in the transition from the
quantum to the classical regime of correlation.

The use of the pulsed high-power laser enables us to tune the PDC
to the high-gain regime while keeping a large pump beam size (of
the order of ${\sim}$1 mm). Thanks to the huge number of radiation
transverse modes, we can concentrate on a portion of the
parametric fluorescence close to the collinear direction and
within a narrow frequency bandwidth around degeneracy. This
portion still contains a large (${>}$1000) number of pairs of
signal/idler correlated phase-conjugate modes, propagating at
symmetrical directions with respect to the pump in order to
fulfill the phase-matching constraints. In the far field, where
the measurement is performed, the couples of modes correspond to
pairs of symmetrical spots, which can be considered as independent
and equivalent spatial replica of the same quantum system. Thanks
to the very large number of these, the statistical ensemble
averaging necessary for the quantum measurement can be solely done
over the \emph{spatial} replicas \emph{for each, single,
pump-laser pulse}. Thus, differently from the experiment in
\cite{Aytur90}, where the statistics was performed over different
temporal replica of the system, here no temporal averages over
successive laser shots are considered. In our experiment the
single-shot measurements reveal sub-shot-noise spatial
correlations for a PDC gain corresponding to the detection of up
to $\simeq$ 100 photoelectrons per mode \cite{Jedrkiewicz04}. In
connection with the work of \cite{Marable02}, where the
correlation between signal/idler phase-conjugate spatial modes was
measured using a band-pass optical parametric amplifier, a
transition from the quantum to the classical regime with
increasing gain was observed. Here we observe a similar
transition, which is attributed to two main effects, one being the
narrowing of the near-field gain profile that occurs at very high
gain in presence of a bell-shaped pump beam. In fact this gain
narrowing effect implies a broadening of the far-field coherence
area, whose transverse size is measured experimentally from the
width of the spatial intensity correlation function evaluated
between the symmetrical signal and idler regions under
consideration. The other effect is the unavoidable inaccuracy in
the experimental determination of the symmetry center of the
signal/idler pattern in the detection plane, and the role of the
CCD pixel size in relation to the coherence area dimension of the
radiation. Our experimental results are in accordance with the
predictions of a 3D quantum model used to describe the
experimental system \cite{Brambilla04}.
\newline\indent
The paper is organized as follows. In section 2 we present the
characteristics of the far field PDC radiation in the pumping
conditions of the experiment, by illustrating the features of the
signal and idler beams detected by the CCD camera at degeneracy.
In section 3 we carefully describe the experimental set-up used
for the spatial correlation measurements and illustrate the
typical single-shot images recorded by the CCD for the evaluation
of the correlation, and in section 4 we present the experimental
results. Our attention is focused on the analysis of the spatial
correlation between signal and idler, to show that the height of
the correlation peak demonstrates the quantum nature of spatial
correlation and implies the apparent violation of a
Cauchy-Schwartz inequality, i.e. a spatial analogue of photon
antibunching in time, as predicted in \cite{Gatti99, Marzoli97,
Lugiato97, Szwaj00}. The numerical simulations showing the
expected transition from the quantum to the classical regime for
increasing gain are presented in section 5, where we also
highlight the important role of the resolution cell size of the
CCD camera (pixel size) and thus, as mentioned above, of the
inaccuracy in the experimental determination of the center of
symmetry of the signal and idler image recorded in the far-field
plane. The conclusions are given in section 6.

\section{Detection of the spatial features of the far-field PDC
radiation by means of the CCD}

Before the quantitative investigation of the existence of spatial
correlations between signal and idler beams, we perform a
preliminary characterization of the generated parametric
radiation. The type II 5x7x4mm${^{3}}$ BBO non-linear crystal,
operated in the regime of parametric amplification of the
vacuum-state fluctuations, is pumped by the third harmonic (352
nm) of a 1ps pulse from a chirped-pulse amplified Nd:glass laser
(TWINKLE, Light Conversion Ltd.). The input and output facets of
the crystal are anti-reflection coated at 352 nm and 704 nm,
respectively. The pump beam (vertically polarized (e)) is
spatially filtered and collimated to a beam waist characterized by
a full width at half maximum (FWHM) of approximately 1 mm at the
crystal input facet, although the fluorescence characterization
has also been performed for smaller beam sizes. The energy of the
352 nm pump pulse can be continuously tuned in the range 0.1-0.4
mJ by means of suitable attenuating filters and by changing the
energy of the 1055nm pump laser pulse, allowing to have a gain $G$
(representing the intensity amplification factor) in the range $10
\le G \le 10^3$. The parametric fluorescence of the horizontally
polarized signal (o) and vertically polarized (e) idler modes is
emitted over two cones, whose apertures depend on the specific
wavelengths (see, e.g., \cite{Berzanskis99, Rubin96}). The BBO
crystal (${\theta=49.05^\circ}$, ${\phi=0}$) is oriented in order
to generate signal and idler radiation cones tangent to the
collinear direction at the degenerate wavelength
$\omega_s=\omega_i=\omega_p/2$ (\emph{s}, \emph{i} and \emph{p}
referring to signal, idler and pump respectively).

\begin{figure}[t]
  \begin{center}
 \includegraphics[width=8.5cm]{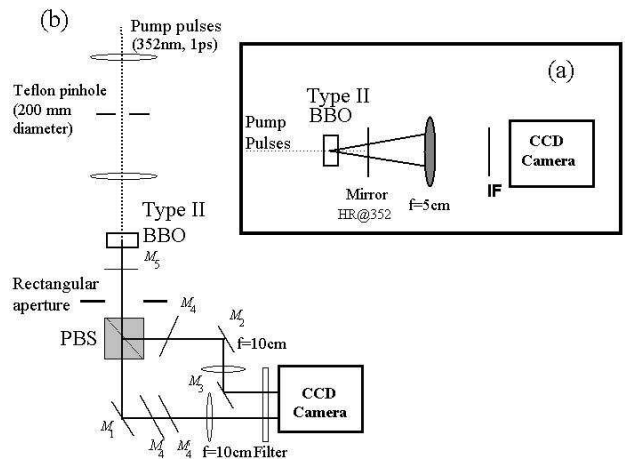}
  \caption{(a) Scheme of the diagnostics for the far-field detection of the degenerate signal and idler ring-type modes.
  (b) Detailed scheme of the experimental set-up used for the spatial correlation measurements. The third harmonic of
   the Nd:Glass laser is used to pump at 352 nm the BBO crystal which is cut for degeneracy at 704nm (${\theta=49.05^\circ}$, ${\phi=0}$).}
   \label{Fig1}
  \end{center}
   \end{figure}

\begin{figure}[t]
\centerline{\includegraphics[width=6cm]{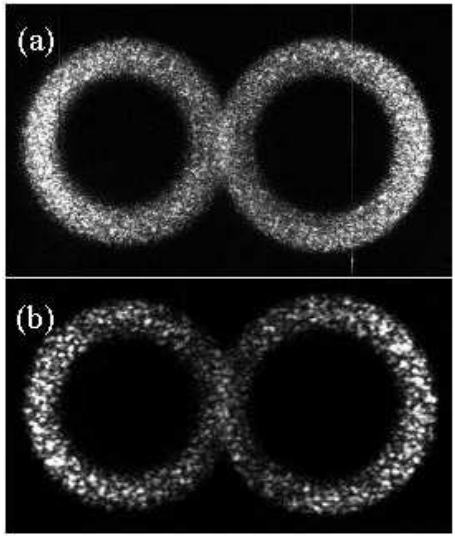}}
   \caption{Experimental far-field images of the degenerate signal (left ring) and idler
    (right ring) beams recorded in single
     shot by the CCD placed in the focal place of a single lens (f=50 cm),
      with pump intensity (a) ${I\approx{30}~{\rm GW/cm}^{2}}$,
     and (b) ${I\approx{50}~{\rm GW/cm}^{2}}$, and FWHM pump beam size of 1 mm (a) and 0.4 mm (b) respectively.}
    \label{Fig2}
  \end{figure}

A simple far-field detection set-up is initially mounted as shown
in Fig.\ \ref{Fig1}(a). A deep-depletion back illuminated charged
coupled device (CCD) camera \cite{Janesick01} (Roper Scientific,
NTE/CCD-400EHRBG1, with quantum efficiency $\eta{\approx89 \%}$ at
704 nm) triggered by a pulse from the laser system, is placed in
the focal plane of a single large-diameter lens (f=5 cm), which
collects at a distance f the far-field PDC radiation emitted by
the BBO. The CCD detection array has 1340 x 400 pixels, with a
pixel size of 20${\mu}$m x 20${\mu}$m. The pump-frequency
contribution is removed by using a normal incidence
high-reflectivity (HR) mirror coated for 352 nm placed after the
BBO. By using a 10-nm broad interferential filter (IF), centered
at 704 nm, we are able to visualize the degenerate signal and
idler far-field beams emitted in the parametric process. It is
worth pointing out that without any spectral filtering, emission
occurs on a very wide range of wavelengths and emission angles
(see, e.g. \cite{Berzanskis99,Rubin96}). Typical far-field images
recorded at degeneracy \emph{in a single shot} (for 1ps pump
pulse) are shown in Fig.\ \ref{Fig2}(a) and Fig.\ \ref{Fig2}(b)
for two different values of the pump intensity and, in the
particular cases illustrated, for two different pump beam sizes.
The ring-shaped angular distribution is determined by the
phase-matching conditions \cite{Berzanskis99}, and the rings at
degeneracy are characterized by an angular width of about
8${^{\circ}}$ each. Note that with this set-up the two rings,
which are emitted along the vertical direction, are recorded by
rotating the CCD by 90${^{\circ}}$ in order to fit the entire ring
pattern inside the rectangular chip.

We draw now the attention on the speckle-pattern aspects of these
images, observing that the dimensions of the speckles or spots
(corresponding to the transverse modes in the far field)
considerably change as a function of the pump beam waist and in
general of the pump beam intensity, and therefore as a function of
the gain. The enlargement of the coherence area of the PDC
radiation in the far field for increasing gain is evident from
Fig.\ \ref{Fig2} and is confirmed by the theory and the numerical
calculations. This effect, which will be extensively analyzed in
the following in sections 5 and 6, turns out to be very important
for the interpretation of the experimental results of the
correlation measurements.

\begin{figure}[t]
\centerline{\includegraphics[width=8cm]{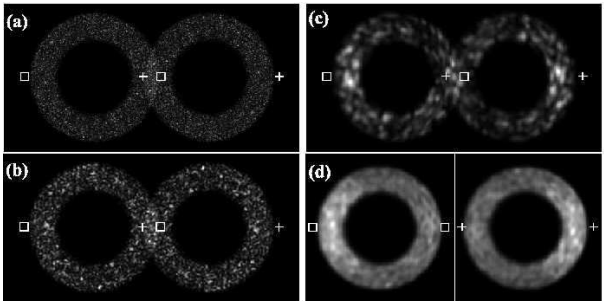}}
   \caption{Numerical far-field images calculated in a single
     shot. (a) and (b) are characterized by the same pump intensities
      as Fig.~\ref{Fig2}(a) and (b) respectively, and
 the pump intensity FWHM is (a) 1 mm and (b) 0.4 mm. (c) has
     pump intensity $I\simeq 100~{\rm GW/cm}^{2}$ and a pump intensity
     FWHM of 0.12 mm. (d) shows an average over 20 shots of the signal
     and idler fields plotted separately for the same numerical conditions of
     (c).  A 16 nm IF was used centered on 704 nm. From the theory in
     Ref.~\cite{Koch}, the predicted hot spot centers for signal
     (idler) are marked with squares (plus). ${\theta=48.90^\circ}$
     was used.}
    \label{Fig3}
\end{figure}

A closer inspection of Fig.~\ref{Fig2} reveals that the photon
distribution on the rings is asymmetrical: there are more photons
along the outer edges of the rings than in the collinear region
where the rings overlap. Moreover, this asymmetry seems to become
stronger in Fig.~\ref{Fig2}(b) which has a narrower pump waist
than Fig.~\ref{Fig2}(a). An explanation of this phenomenon could
be found by considering that for a small pump waist hot spots may
appear in the emission spectrum of PDC. This is due to an
additional phase-matching condition appearing when the Gaussian
shape of the pump is taken into account \cite{Koch}; while always
present when the pump field is Gaussian, it only becomes relevant
when the pump waist is quite small. Instead, for larger pump
waists the usual phase-matching conditions dominate. The hot spots
are predicted to occur along the walk-off direction (in our case
in the $x$ direction in the reference frame of Fig.~\ref{Fig2})
while perpendicular to the walk-off direction less photons should
appear. To investigate if the hot spot phenomenon could explain
the observed asymmetry we present in Fig.~\ref{Fig3}(a)-(b)
numerical simulations for the same parameters as in
Fig.~\ref{Fig2} (for details on the numerics, see
Ref.~\cite{Brambilla04} and section 5). Despite the hot spot
centers being predicted to be located close to the region where
phase matching occurs, no asymmetry in the photon distribution is
observed in Fig.~\ref{Fig3}(a)-(b). In order to see hot spots we
had to reduce the pump waist further, as shown in
Fig.~\ref{Fig3}(c): now hot spots along the walk off direction are
evident. Figure~\ref{Fig3}(d) shows the signal and idler fields
plotted separately after averaging over 20 pump pulses, and a weak
asymmetry between the central (collinear) region and the edges of
the circles is seen.

The numerics suggest that the hot spot phenomenon does occur and
that it might lead to an asymmetry in the photon distribution.
However, the pump waist needed to observe this is almost an order
of magnitude lower than that used in the experiment. Therefore we
checked if imperfections in the pump beam reduce the effective
waist, thus provoking the hot spot effect: simulations including a
Gauss-Hermite pump shape with an overall large pump waist but with
smaller spikes in the profile showed that hot spots could occur
for larger pump waists than those found from a pure Gaussian beam.
Thus a clean pump seems vital as to avoid hot spots from
occurring, in particular considering that in the crystal Kerr
nonlinear effects would amplify any spikes in the pump profile (an
effect we have not accounted for in the numerics). In other words
small-scale focusing of the pump (i.e. filamentation) might occur
or at least be initiated in the crystal due to input-beam
imperfections and Kerr nonlinear response. This might cause the
spectrum to behave as if the actual pump-beam size is smaller than
its FWHM diameter. Nonetheless, we do not believe that hot spots
alone can explain the observed asymmetry. One alternative
explanation might be that absorption in the crystal is saturating
the collinear direction of phase-matching as shown for a type I
optical parametric amplifier by Dou \textit{et al.} \cite{Dou}.
However, this seems an unlikely explanation because the BBO
crystal has a very low absorption, and also Ref.~\cite{Dou}
reports that the saturation occurs for much longer crystals than
our. Preliminary numerical simulations that take into account such
a loss also did not show any qualitative difference with respect
to the data shown in Fig.~\ref{Fig3}, while it remains to be
investigated if nonlinear absorption could lead to more dramatic
effects. We are also suspecting that the interference filter is
creating some unwanted attenuation of the collinear direction, or
distortion effects, and we are currently investigating this in
further detail.

\section{Experimental set-up for spatial correlation
measurements}

The existence of spatial correlation already appears from the
symmetrical properties of the signal and idler patterns recorded
experimentally and shown in Fig.\ \ref{Fig2}. However, to
investigate and reveal the quantum character of the correlation we
use the experimental set-up illustrated in Fig.\ \ref{Fig1}(b)
where, with respect to Fig.\ \ref{Fig1}(a), a different
diagnostics configuration is adopted. We consider a pump beam size
of 1mm and we now select the fluorescence around the collinear
direction by means of a 5mm x 8mm aperture placed 15 cm from the
output facet of the BBO. The radiation is then transmitted through
a polarizing beam splitter (PBS) that separates the signal and
idler beams. The aperture prevents beam clipping by the PBS and
thereby reduces substantially scattered radiation. The beams are
finally sent onto two separate regions of the high efficiency CCD,
which is placed in the common focal plane of the two lenses (f=10
cm) used to image the signal and idler far fields. In contrast to
the case of photon-counting experiments (which rely on coincidence
measurements), and also differently from what was done for the
preliminary characterization of the fluorescence, in this set-up
the correlation measurements are performed without using any
narrow-band IFs, since these unavoidably introduce relevant
transmission losses reducing the visibility of sub-shot-noise
correlations, and could also introduce distortion or even
attenuation effects that, as commented in section 2, are still
unclear. Here the pump-frequency contribution is removed by using
normal incidence (${M_{5}}$) and at 45${^\circ}$ (${M_{4}}$)
high-reflectivity (HR) mirrors coated for 352 nm placed before and
after the PBS, respectively, and a low-band pass colour filter
(90\% transmission around 704 nm) placed in front of the CCD. Note
that a second PBS (not shown in the figure) is placed in the arm
of the (e) idler beam to remove the residual contribution of
ordinary (o) radiation reflected by the first PBS ($3\%$), and a
further HR@352 nm mirror ($M'_{4}$) is placed in the signal arm at
a suitable angle in order to balance the unequal transmission of
radiation in the two arms. All the optical components (except the
colour filter) have anti-reflection coatings at 704 nm. The
estimated quantum efficiency of each detection line, which
accounts for both the transmission losses and the detector
efficiency, is $\eta_{tot}\simeq$75\%.

It is worth pointing out that prior to the experiment for the
detection of quantum spatial correlations, we performed a test of
the capabilities of the scientific CCD camera to perform spatially
resolved measurements of photon shot-noise. The CCD used for the
diagnostic has in fact been calibrated pixel by pixel to
compensate for the gain inhomogeneity of the pixels on the CCD
chip, allowing the retrieval of Poissonian statistics of the
spatial fluctuations of an uniform enlightening in the full range
of the camera dynamics \cite{Jiang03}. The procedure works
efficiently for thermal as well as for laser sources, provided
that the wavelength and the coherence properties of the source are
chosen in order to avoid the formation of equal thickness fringes
in the chip (etaloning effect). Calibration has also allowed the
comparison at the shot-noise level of images recorded at different
places on the chip. At that stage, before the realization of this
quantum correlation experiment, retrieving the shot-noise in the
CCD full dynamic range using classical sources certainly paved the
way to spatially resolved photon noise measurements at the
sub-shot-noise level, and turned out to be a necessary step to
demonstrate quantum properties of images by means of CCD
diagnostic.

\begin{figure}[t]
\centerline{\includegraphics[width=8cm]{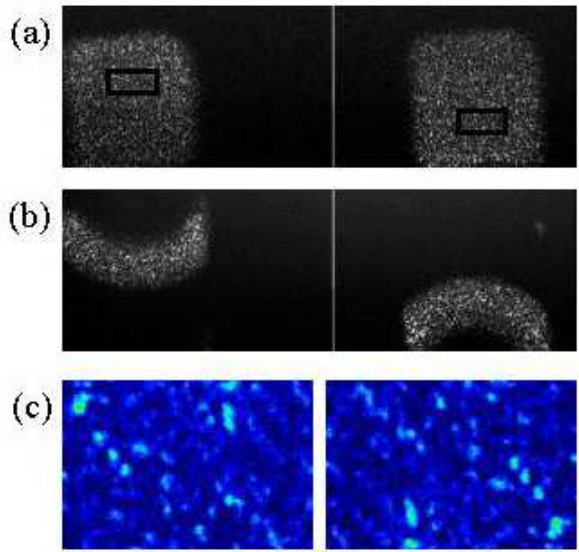}}
 \caption{(a) Single-shot far field image recorded by the CCD for
   a pump intensity $I\simeq 30~{\rm GW/cm}^{2}$. The spatial areas for statistics are delimited by the
    black boxes selected within the degenerate signal and idler modes, spatially localized from the single shot
     image recorded with the 10nm-broad IF (b). (c) Zoom of two symmetrical
     areas of the
     signal and idler far fields.}
    \label{Fig4}
   \end{figure}

Fig.\ \ref{Fig4}(a) shows a typical far-field image recorded in a
single shot in the experimental configuration of Fig.\
\ref{Fig1}(b), where a fairly broadband radiation (\emph{i.e.},
the one transmitted by the rectangular aperture) is acquired in
the signal (left) and idler (right) branches.  The selection of
the desired temporal and angular bandwidth around degeneracy is
made by temporarily inserting in front of the CCD a 10-nm wide IF
around 704 nm, allowing us to locate the collinear degeneracy
point (see Fig.\ \ref{Fig4}(b)). The data analysis is limited
within two rectangular boxes (black frames in Fig.\ref{Fig4}(a))
corresponding to an angular bandwidth of 20 mrad x 8 mrad and to a
temporal bandwidth smaller than 10 nm. The selected regions
contain 4000 pixels each. Since the aim of this work is to
investigate {\it pixel-pair} correlation, and since the size of
the CCD pixel approximately corresponds to the physical size of a
replica (coherence area), the ensemble is large enough to perform
the desired statistics. A zoom of the selected areas is presented
in Fig.\ref{Fig4}(c), where the rather spectacular symmetry of the
intensity distribution in the signal and idler branches shows the
twin-beam character of the phase-conjugate modes.

\section{Detection of quantum spatial correlation:
 Spatial analogue of photon antibunching in time}

\begin{figure}[t]
\centerline{\includegraphics [width=8cm]{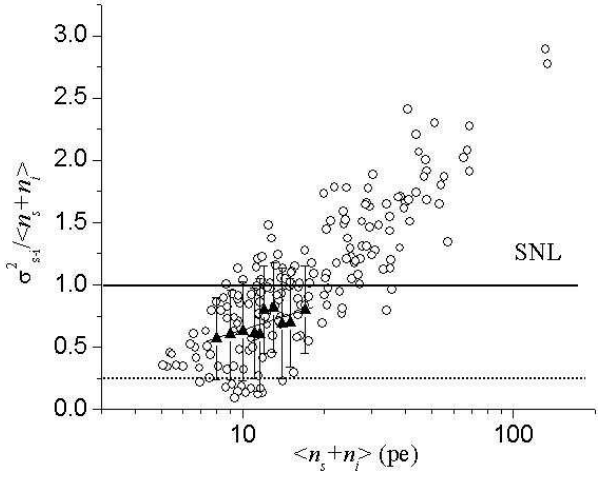}}
    \caption{Intensity difference variance ${\sigma^{2}_{s-i}}$ normalized to the SNL
     ${\langle n_{s}+n_{i} \rangle}$. Each point (white circle)
     corresponds to a single shot measurement where the spatial ensemble
    statistics has been performed over a 100 x 40 pixels region.
     The triangles (each one obtained by averaging the experimental
     points corresponding to a certain gain) and their linear fit illustrate the trend
     of the data in the region between   ${\langle n_{s}+n_{i} \rangle}$ =8 and 20.}
    \label{Fig5}
   \end{figure}

Each of the signal and idler far-field pattern taken separately,
looks like a speckle pattern produced by a pseudo-thermal source,
such as for instance a ground glass illuminated by a laser beam.
When this thermal light is splitted by a macroscopic device as a
beam-splitter, the two resulting beams show a high level of
spatial correlation, which is however limited by shot-noise
\cite{Magatti04, Gatti04a, Gatti04b}. The spatial correlation of
the signal and idler beams generated by PDC is instead of
microscopic origin, and is not limited by shot-noise. The aim of
this experiment is to show the sub-shot-noise nature of the
spatial correlation of the PDC beams. We are first interested in
the symmetrical pixel-pair correlation, which is evaluated
experimentally by measuring the variance ${\sigma^{2}_{s-i}}$ of
the PDC photoelectrons (pe) difference ${n_{s}-n_{i}}$ of the
signal/idler pixel-pair versus the mean total number of
down-converted pe of the pixel-pair. This variance is
\begin{equation}\label{1}
\begin{array}{l}
{\displaystyle \sigma^{2}_{s-i}=\langle
(n_{s}-n_{i})^{2}\rangle-\langle n_{s}-n_{i} \rangle^{2}}
\end{array}
\end{equation}
where the averages are spatial averages performed over all the
symmetrical pixel-pairs contained in the chosen regions. Each
single shot of the laser provides a different ensemble,
characterized by its pixel-pair average pe number ${\langle
n_{s}+n_{i}\rangle}$, in turn related to the parametric gain. In
the experiment, ensembles corresponding to different gains are
obtained by varying the pump-pulse energy. We note that the
read-out noise of the detector, its dark current, and some
unavoidable light scattered from the pump, signal and idler fields
contribute with a non-negligible background noise to the process.
This is taken into account by applying a standard correction
procedure (see for example \cite{Mosset04}), by subtracting the
background fluctuations ${\sigma^{2}_{b}}$ from the
\emph{effectively measured} variance ${\sigma^{2}_{(s+b)-(i+b)}}$
of the total intensity difference
(signal+background)-(idler+background) obtaining
${\sigma^{2}_{s-i}}={\sigma^{2}_{(s+b)-(i+b)}}-2{\sigma^{2}_{b}}$.
This background noise, having a standard deviation of 7 counts
(${\pm0.1}$ from shot to shot, estimated by repeating the
measurement with the same pump-pulse energy) is measured in
presence of pulse illumination over an area of the same size of
the acquisition area and suitably displaced from the directly
illuminated region. The validity of the data correction procedure
and the shot-noise level (SNL) calibration are made by sending in
the set-up (with no crystal) through the PBS, a coherent pulsed
beam (@704nm) linearly polarized at ${45^\circ}$: we have
verified, for different laser energies, that the intensity
difference fluctuations from the two coherent portions of beams
recorded on the CCD lie at the SNL only if this background noise
is taken into account and subtracted from the rough data acquired.
\newline\indent
Fig.\ref{Fig5} shows the experimental results where each point is
associated with a different laser shot. The data are normalized to
the shot-noise level, and their statistical spread accounts for
the background correction. Although the noise on the individual
signal and idler beams is found to be very high and much greater
than their SNL (=${\langle n_{s} \rangle}$ and ${\langle n_{i}
\rangle}$ respectively), we observe an evident sub-shot-noise
pixel pair correlation up to gains characterized by ${\langle
n_{s}+n_{i} \rangle{\approx15-20}}$. Since in that regime the
observed transverse size of the coherence areas (\emph{i.e.} of
the modes) is about 2-4 pixels, this approximately corresponds to
100 pe per mode. The maximum level of noise reduction observed
experimentally agrees with the theoretical limit (dotted line in
Fig.\ \ref{Fig5}) determined by the total losses of the system
(${\sim1-\eta_{tot}}$ \cite{Brambilla04}).

\begin{figure}[t]
\centerline{\includegraphics [width=8cm]{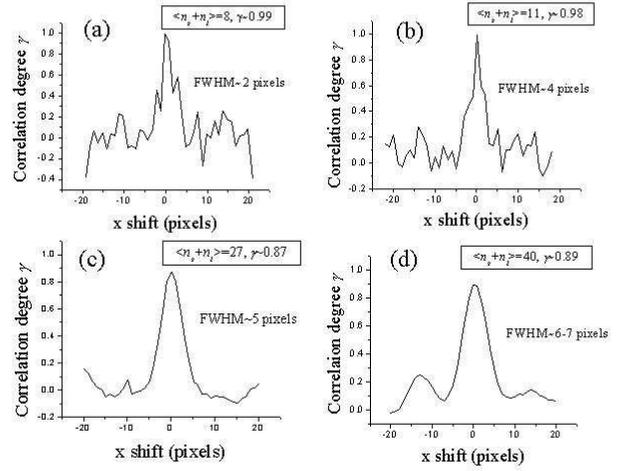}}
  \caption{Correlation degree profiles plotted for four different gain values.}
    \label{Fig6}
   \end{figure}

\begin{figure}[t]
\centerline{\includegraphics [width=8cm]{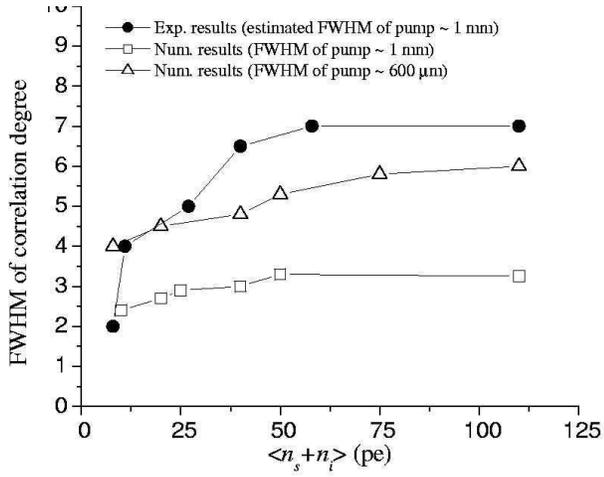}}
    \caption{Experimental and numerical results for the transverse mode size
evaluated from the FWHM of the correlation degree profiles for
different gains.}
    \label{Fig7}
   \end{figure}

We can have an idea of the transverse size of the mode by looking
at the standard two-dimensional cross-correlation degree
\begin{equation}\label{2}
\begin{array}{l}
{\displaystyle \gamma=\frac{\langle n_{s}n_{i} \rangle-\langle
n_{s}\rangle \langle
n_{i}\rangle}{\sqrt{\sigma^{2}_{s}\sigma^{2}_{i}}},}
\end{array}
\end{equation}
between all the angularly symmetrical signal and idler pixels
contained within the black boxes (see Fig.\ \ref{Fig4}). This can
be plotted for instance as a function of the horizontal and
vertical shifts of the recorded image on the CCD, keeping fixed
the position of the boxes. In general ${\mid\gamma\mid\leq1}$ with
${\gamma=1}$ for perfect correlation. The transverse section of
the correlation function $\gamma$ plotted as a function of the
horizontal shift $x$ (in pixel units) and obtained from four
single-shot images corresponding to different gains is presented
in Fig.\ref{Fig6}(a)-(d). We can notice how the FWHM of the curves
increases for increasing gain, clearly reflecting the increment of
the speckles size (and thus of the transverse mode size) already
observed in Fig.\ref{Fig2}. For instance the graph obtained from a
single-shot image characterized by ${\langle n_{s}+n_{i}
\rangle{\approx8}}$ (see Fig.\ref{Fig6}(a)) reveals a transverse
mode size of about 2 pixels, while in the last case plotted in the
figure where ${\langle n_{s}+n_{i} \rangle{\approx40}}$ the FWHM
of the correlation function reaches a dimension of almost 7
pixels. As expected, virtually perfect correlation (in our case we
have peak values of $\gamma$ up to ${\simeq0.99}$) is obtained for
perfect determination (\emph{i.e.} within one pixel) of the center
of symmetry between the signal and the idler regions.
Fig.\ref{Fig7} illustrates the trend (as a function of the gain)
of the coherence area transverse size evaluated from the FWHM of
the correlation degree profiles. The experimental results (here
associated with six images, the same as in Fig.\ref{Fig6} with in
addition two recorded for higher gains) are compared with
numerical data calculated for two different pump size conditions.
Although the qualitative trend between experiment and numerics is
in good agreement, we observe a quantitative discrepancy between
the experimental and numerical data values. This discrepancy is
attributed either to an overestimation of the pump beam size in
the laboratory (measured at a different time from the data
acquisition period -e.g. one month before-, and possibly measured
for different conditions of the laser source, which is very
sensitive to laboratory temperature fluctuations and very
sensitive to the laser regenerative amplifier misalignment); or to
a suspected narrowing of the third-harmonic pump profile for very
high gains that could occur as a consequence of the nonlinear
generation process, since the pump energy increase is performed by
increasing in fact the fundamental laser pulse energy. This is a
point which is currently under further investigation and new pump
beam characterizations are going to be performed.

\begin{figure}[t]
\centerline{
\includegraphics [width=8cm]{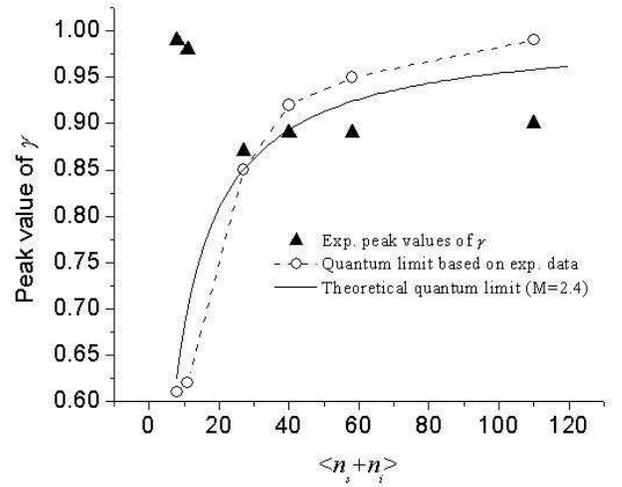}}
   \caption{Experimental correlation degree (triangles) measured from five signal/idler
far-field images for different values of the gain.}
    \label{Fig8}
  \end{figure}

It is interesting to note that the quantum nature of the
correlation can also be estimated from the peak value of the
correlation degree. As a matter of fact, since
\begin{eqnarray}\label{3}
\sigma^{2}_{s-i}=\langle (n_{s}-n_{i}-\langle n_{s}
\rangle-\langle n_{i} \rangle)^{2} \rangle=\nonumber\\
\sigma^{2}_{s}+\sigma^{2}_{i}-2(\langle n_{s}n_{i} \rangle-\langle
n_{s} \rangle \langle n_{i} \rangle)
\end{eqnarray}
the sub-shot noise condition for the intensity difference variance
\begin{equation}\label{4}
\begin{array}{l}
{\displaystyle \sigma^{2}_{s-i}<\langle n_{s}+n_{i} \rangle}
\end{array}
\end{equation}
can be rephrased in the form
\begin{equation}\label{5}
\begin{array}{l}
{\displaystyle \gamma>1-\frac{\langle
n_{s,i}\rangle}{\sigma^{2}_{s,i}}}
\end{array}
\end{equation}
if we use (\ref{2}) and assume that
\begin{equation}\label{6}
\begin{array}{l}
{\displaystyle \langle n_{s}\rangle\simeq\langle
n_{i}\rangle\equiv\langle n_{s,i}\rangle~~ and~~
\sigma^{2}_{s}\simeq\sigma^{2}_{i}\equiv\sigma^{2}_{s,i},}
\end{array}
\end{equation}
as also confirmed experimentally within a good approximation. We
can rewrite (\ref{5}) further by taking into account that, as is
well known, the signal and idler beams taken alone display a
thermal-like statistics. Therefore,
$\sigma^2_{s}\approx\sigma^2_{i}\equiv \sigma^2_{s,i}\approx
\langle n_{s,i} \rangle (1+ \langle n_{s,i} \rangle/M)$
\cite{Goodman00}, where $M$ is the degeneracy factor representing
the number of spatial and temporal modes intercepted by the pixel
detectors. In the condition of the experiment the pump duration is
slightly longer than the PDC coherence time while the pixel area
is smaller than the coherence area, so that $M$ is expected to be
only slightly larger than unity \cite{Mosset04}. Using this,
(\ref{5}) becomes
\begin{equation}\label{7}
\gamma> \frac{\langle n_{s,i}\rangle}{M+\langle n_{s,i}\rangle}.
\end{equation}
Fig.\ref{Fig8} illustrates the trend of the peak values of
$\gamma$ for different gains (black triangles), extracted from the
same images used for Fig.\ref{Fig7}. The dashed-dotted curve
corresponds to the quantum standard limit $\gamma_{lim}$ obtained
by interpolation of the function $\gamma_{lim}=1-{\langle
n_{s,i}\rangle}/{\sigma^{2}_{s,i}}$, calculated for different
gains using the values for the mean and variance obtained from the
experimental far-field patterns considered. The full line is the
theoretical limit obtained from (\ref{7}) by using $M=2.4$ as a
fitting parameter. We thus observe a spatial quantum correlation
whenever the value of $\gamma$ lies in the region above the
theoretical quantum limit. This limit becomes stronger (e.g more
demanding) as the gain increases. The experimental correlation
values obtained are, as expected, compatible with the trend of the
data plotted in Fig.\ref{Fig5}, and highlight a quantum
correlation region up to values of ${\langle n_{s}+n_{i} \rangle}$
of about 20 pe (also corresponding to at least 100 pe per mode).
For instance the first three triangles on the left correspond to
three images characterized by an intensity difference variance
that is clearly below the SNL in Fig.\ref{Fig5}, while the other
three triangles correspond to images that are characterized by an
intensity difference variance above the SNL. Similarly to the case
of the points plotted in Fig.\ref{Fig5}, the background noise
correction leads to a statistical spread of the measured data also
for the correlation degree. In relation to this comment we mention
then that the experimental points (black triangles) represented in
Fig.\ref{Fig8} have been extracted from images characterized, for
a given gain, by the highest correlation found.

Finally we note that if we multiply (\ref{5}) by
${\sigma^{2}_{s-i}}$, with the help of (\ref{2}) and (\ref{6}) we
obtain
\begin{equation}\label{8}
\langle n_{s}n_{i}\rangle-\langle n_{s}\rangle\langle
n_{i}\rangle>\sigma^{2}_{s,i}-\langle n_{s,i}\rangle
\end{equation}
i.e., by defining
\begin{equation}\label{9}
\langle \delta n_{s}\delta n_{i}\rangle\equiv\langle
n_{s}n_{i}\rangle-\langle n_{s}\rangle\langle n_{i}\rangle,
\end{equation}
and
\begin{equation}\label{10}
\langle:\delta n^{2}_{s,i}:\rangle\equiv\sigma^{2}_{s,i}-\langle
n_{s,i}\rangle,
\end{equation}
where the symbol : : indicates the normal ordering, we find the
following inequality condition
\begin{equation}\label{11}
\langle \delta n_{s}\delta n_{i}\rangle>\langle:\delta
n^{2}_{s,i}:\rangle=[\langle:\delta
n^{2}_{s}:\rangle\langle:\delta n^{2}_{i}:\rangle]^{1/2},
\end{equation}
equivalent to the sub-shot-noise level condition (\ref{5}), and
which states that the cross-correlation between signal and idler
is larger than the (normally ordered) self-correlation. This
corresponds to an apparent violation of the Cauchy-Schwartz
inequality. This effect, which was predicted in \cite{Marzoli97,
Lugiato97, Szwaj00} for the case of the optical parametric
oscillator and then generalized \cite{Gatti99} to the case of the
travelling-wave optical parametric amplifier, represents a spatial
analogue of the phenomenon of photon antibunching in time.

\section{Transition from quantum to classical regime: numerical
results}

\begin{figure}[t]
\centerline{
\includegraphics[width=8cm]{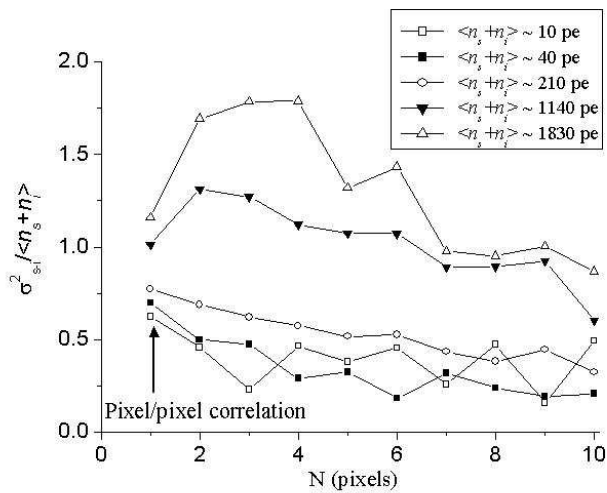}}
   \caption{Numerical calculation of ${\sigma^{2}_{s-i}}$ (normalized to SNL)
     between symmetrical portions of signal and idler plotted as a
     function of the detection area represented by N x N binned
     pixels. Different curves correspond to different values of the
     gain characterized by the mean number of down-converted pe per
     pixel pair ${\langle n_{s}+n_{i} \rangle}$.}
     \label{Fig9}
\end{figure}

In order to interpret the observed transition from quantum to
classical regime in Fig.\ \ref{Fig5}, we present in Fig.\
\ref{Fig9} the results of the numerical calculations. The full
quantum model accounts for the two transverse and the temporal
degrees of freedom with propagation along the crystal, for the
angular and chromatic material dispersion up to the second order,
for the finite spatial and temporal widths of the Gaussian pump
pulse, and for the experimental quantum efficiency ${\eta_{tot}}$.
It has been important to investigate how these features affect the
spatial quantum correlation phenomena initially predicted by the
plane wave pump theory, also to identify the conditions under
which they can be observed in the experiment. The quantum
description of the PDC is performed by treating the pump as a
known classical field which propagates linearly inside the
crystal, while the down-converted fields are quantized. The
quantum averages needed in the calculations are evaluated through
a stochastic method based on Wigner representation, as described
in \cite{Brambilla04}.

Fig.\ \ref{Fig9} presents ${\sigma^{2}_{s-i}}$, normalized to the
SNL, \emph{vs.} the size of the detection area for different
gains. Each point is the result of a statistics performed over one
single laser-shot. The case N=1 corresponds to the experiment. The
simulations (data not shown) outline that, in spite of the fixed
pump-beam diameter, the signal and idler beam diameters at the
crystal output strongly depend on the gain and decrease when the
latter increases. This can be easily interpreted when considering
that the signal and idler beam size maps not the pump-beam profile
but the actual parametric amplification gain profile
${G(\textbf{r})\sim \rm cosh^{2}[{\sigma}A(\textbf{r})L]}$
\cite{Akhmanov} (L being the crystal length, A the pump field
amplitude and ${\sigma}$ a parameter proportional to the setting
characteristics), as long as filtering due to the limited spatial
bandwidth does not take place \cite{Ditrapani98}. On narrowing the
size of the PDC beams, the coherence areas in the far field
(\emph{i.e.} the modes) increase their size, as straightforward
consequence of the convolution theorem in Fourier analysis \cite
{Berzanskis99}. This is precisely the effect observed
experimentally and revealed either directly by looking at the
far-field patterns recorded by the CCD, or by analyzing the
correlation degree function profile as discussed in section 4.
Since revealing quantum correlations requires detection areas
larger (or comparable) to the mode size (as also discussed in
\cite{Brambilla04}), it is necessary when increasing the gain to
have larger detectors in order to obtain sub-shot-noise variance
as shown in Fig.\ \ref{Fig9}. Note that Fig.\ \ref{Fig9} evidences
the transition from quantum to classical regime in case of
single-pixel detection (N=1) for a gain that is higher than in the
experiment. Indeed, in the experiment, excess noise is observed
already for ${\langle n_{s}+n_{i} \rangle}>$20, which we attribute
first to the effect of residual scattered light whose contribution
grows linearly with the radiation fluence and is thus expected to
overcome the shot noise at large pumping, and second to the
uncertainty in the determination of the symmetry center of the
signal and idler image portions.

In fact although the increase of the coherence area with the gain
leads to a deterioration of the signal-idler correlation,
numerical calculations have shown that this phenomenon alone is
not sufficient to explain the steepness of the slope given by the
ratio $\sgmsnl\equiv\sigma_{s-i}^2/\langle n_s+n_s\rangle$ as a
function of the shot-noise $\langle n_s+n_s\rangle$ displayed in
the plot of Fig.\ \ref{Fig5}.  As we shall now show, another
important feature which contributes to this behaviour is related
to the inaccuracy in the determination of the center of symmetry
of the far-field pattern in the detection plane. In our experiment
the typical error $\Dx$ between the selected and the actual center
of symmetry is on the order of half the size of the pixels of the
CCD, that is about 10~$\mu$m. $\Dx$ can therefore be a significant
fraction of the far-field coherence length, $x_{coh}$ (transverse
size of the spatial mode), depending on the gain values considered
in the experiment (here $x_{coh}$ is typically on the order of
50~$\mu$m). We are therefore in an intermediate situation between
the ideal condition where $\Dx\ll x_{coh}$ -- for which the theory
predicts sub-shot noise correlation -- and the opposite limit
$\Dx\gg x_{coh}$ in which the selected pixel pairs used to perform
the statistics are completely uncorrelated. In this latter case
the variance of $n_s-n_i$ is simply the sum of the variance of
$n_s$ and $n_i$ (since $\langle n_s n_i \rangle=\langle n_s
\rangle \langle n_i \rangle$), that is
$\sigma^2_{s-i}=\sigma^2_{s}+\sigma^2_{i}$.  As mentioned before,
since the signal and idler beams taken alone display thermal
statistics, $\sigma^2_{s}\approx\sigma^2_{i}\equiv
\sigma^2_{s,i}\approx \langle n_{s,i} \rangle (1+ \langle n_{s,i}
\rangle/M)$, having assumed $\langle n_s \rangle\approx \langle
n_i \rangle\equiv \langle n_{s,i} \rangle$. The saturation value
of $\sgmsnl$ in the $\Dx\gg x_{coh}$ limit is therefore given by
$\sgmsnlsat=(\sigma^2_{s}+\sigma^2_{i})/\langle
n_s+n_i\rangle\approx 1+\langle n_{s,i}\rangle/M$.

\begin{figure}[t]
\centerline{
\includegraphics[width=8cm]{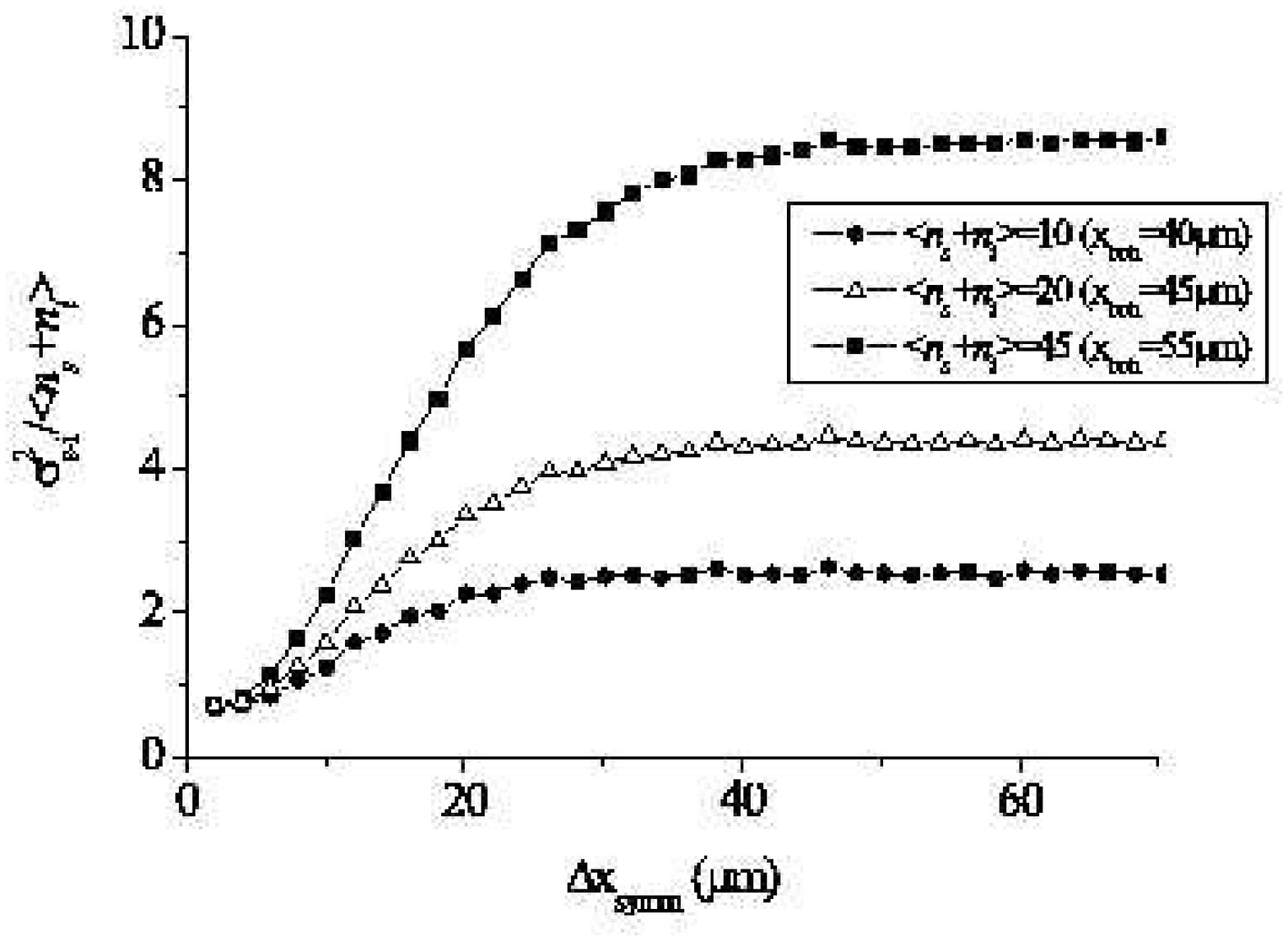}}
\caption{Plot of $\sgmsnl\equiv\sigma_{s-i}^2/\langle
n_s+n_i\rangle$ as a
  function of $\Dx$ for different value of the gain.  For $\Dx>x_{coh}$
  $\sgmsnl$ saturates to a value on the order of $1+\langle
  n_{s}+n_i\rangle/(2M)$, from which $M\approx 3$ is deduced. As a raw
  approximation the transition from sub-shot-noise to above shot-noise
  arises for $\Dx/x_{coh}\propto 1/\sgmsnlsat\approx (1+\langle
  n_{s,i}\rangle/M)^{-1} $.}
\label{Fig10}
\end{figure}

The general behaviour for arbitrary values of the $\Dx$ parameter
is illustrated in Fig.\ \ref{Fig10}, where $\sgmsnl$ is obtained
from numerical simulations by varying $\Dx$ in the $x$-axis
(walk-off) direction for different gain values.  As $\Dx$ becomes
larger than $x_{coh}$ the data saturate, and from these saturation
values we estimate the degeneracy factor $M\approx 3$ by comparing
them to the predicted values $\sgmsnlsat=1+\langle n_{s,i}
\rangle/M$. This $M$ value is quite close to the one estimated
experimentally ($M=2.4$) in the previous section. From the
behaviour of these curves it can be inferred that the transition
from sub-shot-noise to above-shot-noise correlations occurs
approximately when $\Dx/x_{coh}$ exceeds a critical value $\propto
1/\sgmsnlsat\approx (1+\langle n_{s,i}\rangle/M)^{-1} $.

\begin{figure}[t]
\centerline{
\includegraphics[width=8cm]{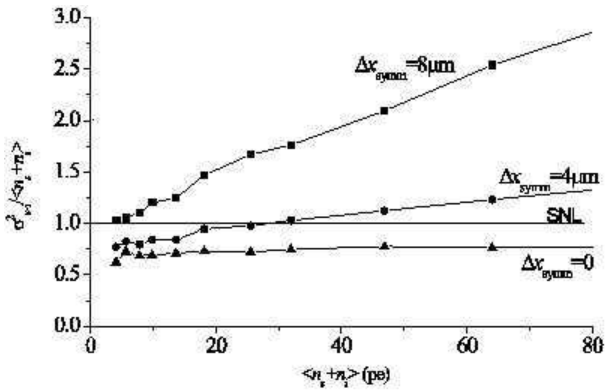}}
\caption{
  Plot of $\sgmsnl\equiv\sigma_{s-i}^2/\langle n_s+n_i\rangle$ as a
  function of $\langle n_s+n_i\rangle$ for different value of $\Dx$.
  The slope of the curves is on the order of $(2M)^{-1}
  \Dx/x_{coh}$}
\label{Fig11}
\end{figure}

It also follows from Fig.\ \ref{Fig10} that $\sgmsnl$ increases
almost linearly with $\langle n_s+n_i\rangle$ with a slope which
is on the order of $(2M)^{-1} \Dx/x_{coh}$. The plot of Fig.\
\ref{Fig11} confirms this result, showing how rapidly the
sub-shot-noise correlation is lost as $\Dx$ becomes a significant
fraction of $x_{coh}$. We note that in the range of $\langle
n_s+n_i\rangle$ considered in those simulations (which are those
of the experiment) the effect of the broadening of the coherence
area due to the increasing gain is negligible (see also the
numerical resuls of Fig.\ \ref{Fig9} in the case N=1), as can be
seen from the $\Dx=0$ curve which is almost horizontal. For high
values of the gain, the inaccuracy in the determination of the
symmetry center becomes therefore particularly relevant and its
contribution to the loss of correlation is expected to exceed the
contribution due to broadening of the coherence area. Other
contributions not included in the numerical model, such as light
scattering from the environment, becomes also important as the
number of PDC photons increases.

\section{Conclusions}

In conclusion, we have presented experimental and numerical
results demonstrating the quantum spatial features of the
radiation generated in parametric-down-conversion in the high-gain
regime and measured by means of a scientific CCD camera. We have
first illustrated the speckle-like patterns of the fluorescence
detected in particular at degeneracy, pointing out the effect of
coherence area enlargement for increasing gain. The experimental
investigation of the quantum aspects of the intensity correlations
between signal and idler leads to the conclusion that twin beams
of light generated in PDC exhibit sub-shot-noise spatial
correlation. This has been shown by measuring in the far-field an
evident quantum noise reduction on the signal/idler intensity
difference, and by having high peak correlation degree values
lying above the standard quantum limit. The latter phenomenon
corresponds to an apparent violation of the Cauchy-Schwartz
inequality, i.e. to the spatial analogue of photon antibunching in
time, which was predicted in \cite{Gatti99, Marzoli97, Lugiato97,
Szwaj00}. A transition to above shot-noise correlation is observed
as the gain increases. The theory and numerical simulations show
how a quantum-to-classical transition is expected to occur because
of a narrowing of the signal/idler beams with increased gain. This
leads in turn to a larger far-field mode size and therefore also
to the need of larger pixels to observe sub-shot-noise correlation
\cite{Brambilla04}. However numerical calculations have also shown
that this phenomenon alone is not sufficient to explain the
steepness of the slope of the transition observed experimentally.
In addition to the effect of residual scattered light, whose
contribution grows linearly with the radiation fluence and is thus
expected to overcome the shot noise at large pumping, an important
feature contributing to this behavior is found to be related to
the inaccuracy in the determination of the center of symmetry of
the signal/idler recorded pattern in the far-field plane. The
importance of the results presented in this work lies in the fact
that this is the first experimental investigation of quantum
spatial correlations in the high gain regime, where the huge
number of transverse spatial modes is detected in single shot by
means of a high-quantum-efficiency CCD.
\newline
\newline
\textbf{Acknowledgements}

This work was supported by projects FET QUANTIM "Quantum imaging",
PRIN of MIUR "Novel devices based on quantum entanglement", FIRB01
"Space-time phenomena in nonlinear optics", INTAS 2001-2097. M.B.
acknowledges support from the Carlsberg Foundation. We thank
Yunkun Jiang for useful discussions and many helpful advices.


\end{document}